\documentclass[10pt,conference]{IEEEtran}
\usepackage{amsmath,amsfonts,amsthm}
\usepackage{algorithm,algorithmic}
\usepackage{xspace}
\usepackage{booktabs}
\usepackage{graphicx}
\usepackage{textcomp}
\usepackage[table,dvipsnames]{xcolor}
\usepackage{listings}
\usepackage{multirow}
\usepackage{relsize}
\usepackage{pifont}
\usepackage{hyperref}
\usepackage{paralist}
\usepackage{enumitem}
\usepackage{xurl}
\usepackage{tabularx}   
\usepackage[flushleft]{threeparttable}
\usepackage{bm}
\usepackage{svg}
\usepackage{tikz}
\usepackage[most]{tcolorbox}

\usepackage{tikz}

\def\BibTeX{{\rm B\kern-.05em{\sc i\kern-.025em b}\kern-.08em
    T\kern-.1667em\lower.7ex\hbox{E}\kern-.125emX}}

\definecolor{mycolor}{rgb}{0.122, 0.435, 0.698}
\definecolor{gray1}{gray}{0.3}

\definecolor{codegreen}{rgb}{0,0.6,0}
\definecolor{codegray}{rgb}{0.5,0.5,0.5}
\definecolor{codepurple}{rgb}{0.58,0,0.82}
\definecolor{blackcolour}{rgb}{0.95,0.95,0.92}
\lstdefinestyle{mystyle}{
    commentstyle=\color{codegreen},
    keywordstyle=\color{magenta},
    numberstyle=\tiny\color{codegray},
    stringstyle=\color{codepurple},
    basicstyle=\tiny\ttfamily,
    breakatwhitespace=false,
    breaklines=true,
    captionpos=b,
    keepspaces=true,
    numbers=left,
    numbersep=5pt,
    showspaces=false,
    showstringspaces=false,
    showtabs=false,
    tabsize=2,
    columns=fixed
}
\usepackage{comment}
\usepackage{pifont}
\usepackage{pbox}
 \usepackage[flushleft]{threeparttable}

\usepackage{booktabs}

\usepackage{diagbox}
\usepackage{tablefootnote}

\usepackage{amsmath}
\usepackage{caption}

\usepackage{subcaption}

\usepackage{hyperref}
\usepackage{adjustbox}
\usepackage{multirow}
\usepackage{tabularx}
\usepackage{enumitem}
\usepackage{framed}
\usepackage[linesnumbered,ruled,algo2e]{algorithm2e}
\SetKwInput{KwInput}{Input} \SetKwInput{KwOutput}{Output} \SetKwInput{KwProcedure}{Procedure}
 \SetCommentSty{mycommfont}
\usepackage{graphicx}
\usepackage{textcomp}
\usepackage{xcolor}
\usepackage{multicol}
\usepackage{balance}
\usepackage{subcaption}
\usepackage{epstopdf}
 \AppendGraphicsExtensions{.pdf}
\usepackage{array}
\usepackage{listings}
\definecolor{Gray}{gray}{0.9}
\definecolor{pgreen}{rgb}{0,0.5,0}
\usepackage{amsthm}
\makeatletter
\def\th@plain{%
  \thm@notefont{}
  \itshape 
}
\def\th@definition{%
  \thm@notefont{}
  \normalfont 
} \makeatother
\usepackage{capt-of}

\usepackage{color, colortbl}
\definecolor{grey}{rgb}{0.7,0.7,0.7}

\newcommand{\lstbg}[3][0pt]{{\fboxsep#1\colorbox{#2}{\strut #3}}}
\lstdefinelanguage{diff}{
  basicstyle=\ttfamily\scriptsize,,
  morecomment=[f][\lstbg{red!20}]-,
  morecomment=[f][\lstbg{green!20}]+,
  morecomment=[f][\lstbg{yellow!20}]++,
  morecomment=[f][\textit]{@@},
  texcl=false
}

\usepackage{tcolorbox}
\usepackage{xcolor}
\definecolor{indiagreen}{rgb}{0.07, 0.53, 0.03}
\definecolor{hycolor}{rgb}{0.7,0.7,0.3}
\definecolor{darkbrown}{rgb}{0.4, 0.26, 0.13}

\usepackage{listings}
\usepackage{xcolor}
\definecolor{main-color}{rgb}{0.6627, 0.7176, 0.7764}
\definecolor{string-color}{rgb}{0.3333, 0.5254, 0.345}
\definecolor{key-color}{rgb}{0.8, 0.47, 0.196}

\lstdefinestyle{mystyle} {
    language = Java,
    basicstyle = {\ttfamily \color{main-color}},
    stringstyle = {\color{string-color}},
    keywordstyle = {\color{key-color}},
    keywordstyle = [2]{\color{lime}},
    keywordstyle = [3]{\color{yellow}},
    keywordstyle = [4]{\color{teal}},
    morekeywords = [3]{<<, >>},
    morekeywords = [4]{++},
    basicstyle=\ttfamily\scriptsize,
    commentstyle=\color{blue}\ttfamily,
    morecomment=[f][\lstbg{red!20}]-,
    morecomment=[f][\lstbg{green!20}]+,
    morecomment=[f][\lstbg{yellow!20}]++,
    morecomment=[f][\lstbg{yellow!20}]--,
    morecomment=[f][\textit]{@@},
    breaklines=false,
    texcl=false
}

\lstnewenvironment{bash}[1][]
  {\lstset{language=[latex]TeX}\lstset{%
   numbers=left,numberstyle=\scriptsize,
   commentstyle=\color{pgreen},
   xleftmargin=5pt,
   xrightmargin=5pt,
   aboveskip=\bigskipamount,
   belowskip=\bigskipamount, #1,
}} {}

\lstdefinestyle{testlstcolor}{
    language={sh},
    moredelim=**[is][\color{red}]{~}{~},
    moredelim=**[is][\color{blue}]{<}{>},
    moredelim=**[is][\bfseries]{***}{***},
    moredelim=**[is][\color{green}]{~~}{~~},
    showstringspaces=false,
    basicstyle=\ttfamily,
    literate={\\~}{{\textasciitilde}}1
        {\\<}{{\unichar{"003C}}}1
        {\\>}{{\unichar{"003E}}}1
}

\newcolumntype{L}[1]{>{\raggedright\let\newline\\\arraybackslash\hspace{0pt}}m{#1}}
\usepackage{tikz}

\usepackage{xspace}

\definecolor{darkgreen}{rgb}{0.0, 0.5, 0.0}
\definecolor{darkred}{rgb}{0.82, 0.1, 0.26}

\usepackage[dvipsnames]{xcolor}

\newcommand{\allrepositories}{461\xspace}
\newcommand{\allprojects}{\allrepositories}

\definecolor{findingback}{RGB}{235,241,249}     
\definecolor{findingborder}{RGB}{70,130,195}
\newtcolorbox{finding}[1][]{%
  enhanced,
  colback=findingback, colbacktitle=findingback, colframe=white,
  coltitle=black, fonttitle=\bfseries,
  borderline west={3pt}{0pt}{findingborder},
  boxrule=0pt, titlerule=0pt, arc=0pt, sharp corners,
  left=2mm, right=2mm, top=0.8mm, bottom=0.8mm,
  boxsep=0.6mm, before skip=5pt, after skip=5pt,
  title={#1},
}

\newcommand{\code}[1]{\texttt{#1}}
\newcommand{\niscan}{\textsc{NIScan}}
\newcommand{\ini}{\textsc{INI}}

\begin{document}
\title{
From Codebases to LLMs: Non-Inclusive Naming in Linux Foundation Repositories
}

\author{
  \IEEEauthorblockN{Honghao Tan\IEEEauthorrefmark{1},
    Md Nafiu Rahman\IEEEauthorrefmark{2},
    Shin Hwei Tan\IEEEauthorrefmark{1}}
  \IEEEauthorblockA{\IEEEauthorrefmark{1}Concordia University, Montreal, Canada\\
    honghao.tan@mail.concordia.ca, shinhwei.tan@concordia.ca}
  \IEEEauthorblockA{\IEEEauthorrefmark{2}Brac University, Dhaka, Bangladesh\\
    nafiu.rahman@bracu.ac.bd}
}

\maketitle
\bstctlcite{IEEEexample:BSTcontrol}

\begin{abstract}

Since 2020, the Linux Foundation and the multi-organization Inclusive Naming Initiative (\ini) have encouraged open-source projects to replace non-inclusive terms such as \code{master}/\code{slave} and \code{whitelist}/\code{blacklist}. Although these recommendations have been widely adopted, there is limited empirical evidence on their long-term adoption across Linux Foundation (LF) projects or their implications for AI-assisted software development. In this paper, we present \niscan{}, a multilingual static-analysis framework that detects non-inclusive terminology across source code and related software artifacts using the INI vocabulary. Using \niscan{}, we conduct the first ecosystem-scale study of inclusive naming across \allrepositories\ Linux Foundation repositories. Our analysis shows that non-inclusive terminology has declined by approximately 47\% since 2020, yet adoption remains incomplete: 62.7\% of repositories still contain at least one Tier-1 non-inclusive identifier, while most remaining terminology resides outside source code in documentation, comments, configuration files, and other software artifacts. We further show that repository size, programming language, project functionality, and ecosystem
are stronger predictors of term inclusiveness in LF repositories rather than foundation governance. To examine the implications for AI-assisted software development, we conduct a case study evaluating whether large language models (LLMs) can reconstruct legacy non-inclusive identifiers from surrounding program context. The results show that historical naming decisions remain embedded in model predictions even after identifiers have been renamed. Overall, our study findings provide the first ecosystem-scale assessment of inclusive naming adoption within the Linux Foundation and highlight the importance of addressing terminology residue to support responsible naming and ethically sourced code generation.
\end{abstract}

\vspace{-3mm}
\section{Introduction}
\label{sec:intro}

Inclusive naming has become an important software engineering practice for improving the accessibility, professionalism, and inclusiveness of software systems~\cite{hyrynsalmi2025making}. In response to growing recognition that terms such as \emph{master/slave}, \emph{whitelist/blacklist}, and other historically established identifiers may be perceived as non-inclusive or offensive, software communities initiated one of the largest coordinated terminology migrations in the history of open-source software. In 2020, GitHub changed the default branch name for new repositories from \emph{master} to \emph{main}, while Linux Foundation (LF), together with the Cloud Native Computing Foundation, IBM, Red Hat, Cisco, and VMware, co-founded the Inclusive Naming Initiative (INI)~\cite{ini2024wordlist} to promote the adoption of more inclusive terminology across software systems. The movement has since expanded beyond community-driven recommendations. On 19 June 2025, the IEEE Standards Association approved IEEE Std 3400-2025~\cite{ieeestandard}, providing the first international standard for evaluating and adopting inclusive terminology in engineering systems. These industry and standards efforts show that inclusive naming is now recognized as an important software engineering practice rather than merely a language preference.

Despite coordinated efforts by industry and open-source communities, inclusive naming remains an ongoing software engineering challenge rather than a completed transition. While inclusive naming guidelines have been widely adopted by organizations and projects, there is limited empirical evidence on their long-term adoption across the Linux Foundation ecosystems. Five years after the Inclusive Naming Initiative (INI) was introduced, it remains unclear \emph{to what extent non-inclusive terminology persists across Linux Foundation (LF) software projects}, where such terminology appears, and \emph{whether the recommended naming practices have been widely adopted} in practice.

Ensuring inclusive naming is important because names are unusually persistent software artifacts. Once embedded in identifiers, APIs, configuration files, documentation, and operational workflows, terminology can outlive the code that introduced it and become increasingly costly to change. The continued presence of non-inclusive terminology constitutes a form of \emph{linguistic technical debt}~\cite{peruma2022refactoring,mrad2026effectiveness}, affecting not only source code but also developer documentation, user interfaces, and project governance. As software development practices continue to evolve, the consequences of this technical debt extend beyond maintaining existing systems to shaping how future software is developed. In our case study with four LLMs, we build upon our empirical results to evaluate modern LLM propagation loops of this linguistic technical debt.

The importance of responsible identifier selection has gained renewed attention in the era of large language models (LLMs), as residual non-inclusive terminology within legacy software repositories is actively ingested, regenerated, and potentially amplified by generative AI.
Prior work shows that code generation models frequently reproduce latent social biases~\cite{huang2025bias} and harmful terminology~\cite{tan2025harmfulness,tan2026dr} embedded within their pre-training distributions.
Building on this line of work, we conduct a preliminary case study in which function names are masked and regenerated using state-of-the-art code LLMs. We observe that these models frequently propose non-inclusive identifiers even when inclusive alternatives are readily available. This suggests that historical naming decisions embedded in software artifacts can persist through model behavior, leading to the reintroduction of non-inclusive terminology into newly generated code. Such a feedback loop highlights a broader concern in AI-assisted software development, where linguistic technical debt may be continuously propagated through model outputs, thereby challenging emerging efforts toward ethically sourced code generation practices~\cite{xu2026makescodegenerationethically}.

Despite growing attention from Linux Foundation (LF), the long-term adoption of inclusive naming remain insufficiently understood. Although tools have been proposed to detect and refactor non-inclusive terminology~\cite{winchester2023harmful,todd2024githubinclusifier}, there is still limited empirical evidence on the prevalence of non-inclusive terminology across LF projects. In particular, it remains unclear where such terminology persists (e.g., source code, documentation, comments, or configuration files), how its usage has evolved since the 2020 ecosystem-wide renaming initiative, and whether project governance influences its removal.

To answer these questions, we present \textbf{\niscan{}}, a multi-language static-analysis tool that measures non-inclusive terminology across source code and related software artifacts. Using \niscan{}, we conduct the first ecosystem-scale study of inclusive naming across \allrepositories Linux Foundation repositories.
Our study makes the following contributions:
\begin{itemize}[leftmargin=*]
    \item \textbf{Framework:} We design and implement \niscan{}, 
    a multilingual static analysis framework that automatically identifies non-inclusive identifiers matching the authoritative Inclusive Naming Initiative (INI) vocabulary, supporting 10 programming language. Our framework achieve a validated overall precision of $95.2\%$ and $100\%$ precision on Tier-1 identifier terms. 
    \item \textbf{Study of Term Inclusiveness in LF repos:} We conduct the first comprehensive study of term inclusiveness across \allrepositories{} actively maintained Linux Foundation (LF) repositories spanning 12 independent sub-foundations, mapping out a corpus of 133,541 matched term occurrences. Our study identify the technical and organizational factors associated with non-inclusive terminology persistence, including project maturity, repository size, primary language, and functionality domain.
    \item \textbf{Case Study on LLM Reconstruction of Non-Inclusive Terms:} We conduct a controlled probe case study across four LLMs (GPT-5.5, Claude Opus 4.8, Gemini 3.1 Pro, and DeepSeek V4 Pro) to evaluate the downstream consequences of non-inclusive terms. We demonstrate that while models heavily prefer inclusive alternatives, historical terminology continues to persist latently within model distributions, resulting in top-3 reconstruction rates between $61\%$ and $83\%$ from surrounding code context. 
\end{itemize}

Our findings reveal that non-inclusive terminology remains widespread
despite years of ecosystem-wide guidance, with non-inclusive terms
persisting far beyond source code into documentation and other project
artifacts. Although term density has declined since 2019, the
reduction has plateaued, and frontier language models still reproduce
non-inclusive identifiers from contextual cues alone. These results
suggest that inclusive naming should be viewed not only as a governance
concern, but also as a software-evolution and AI-software-engineering
challenge.

\section{Background and Related Work}
\label{sec:background}

\subsection{Inclusive-naming Guidance and Its Renewed Relevance}
\label{sec:bg:policy}

The push for inclusive naming crystallised in 2020--2021: GitHub moved new
repositories from \code{master} to \code{main}~\cite{github2020main}, and the
Inclusive Naming Initiative (\ini{}) was established under the Linux
Foundation (§\ref{sec:intro}), publishing a four-tier word list from
``adopt immediately'' to ``no change''~\cite{ini2024wordlist}. The concern is
neither settled nor dated: adoption is non-monotonic (NIST withdrew its own
guidance~\cite{nistir8366}) yet \ini{} stays active (resources updated through
2024--2025, membership expanded), the Linux kernel removed its
long-standing \code{d\_genocide()} function for the same
reason~\cite{linuxkernel2024genocide}, and the
generative-AI era renews it: a 2024--2026 wave documents social bias and
harmful-term generation in code
LLMs~\cite{huang2025bias,du2025faircoder,tan2025harmfulness} that
regenerate the very residue we measure. Yet prior writing on compliance
is limited to position papers, single-project case studies, and surveys; none
gives a quantitative baseline at ecosystem scale, and measuring an adoption
trajectory needs the multi-year hindsight only now available.

\subsection{The Linux Foundation Ecosystem}
\label{sec:bg:lf}
The Linux Foundation (LF) is an umbrella organization whose \emph{sub-foundations} (including CNCF, LF~Energy, FINOS, Hyperledger, and others) each govern an independent set of projects. Some of these sub-foundations publish a \code{landscape.yml}~\cite{cncf2025landscape}, a YAML file that enumerates \emph{member projects} and maps each project to one or more official GitHub \emph{repositories}. As single project may be implemented across multiple repositories (for example, to separate components such as core services, client libraries, and deployment tooling), we use a repository rather than the project as the unit of analysis. Our corpus includes all twelve sub-foundations that publish such landscape files. Several sub-foundations further classify projects based on \emph{maturity}. For example, sub-foundations such as CNCF defines a maturity ladder with three levels: (1) \emph{sandbox} (early-stage), (2) \emph{incubating} (in-progress), and (3) \emph{graduated} (production-ready).

\subsection{Research on Inclusive Naming and Governance}
\label{sec:bg:related}

Identifier naming is known to affect software quality and
comprehension~\cite{deissenbock2005concise,lawrie2007effective,butler2010nameconsistency,binkley2009camelcase,hofmeister2017shorter,allamanis2018survey,arnaoudova2016antipatterns,fakhoury2018cognitive},
but that work treats naming as a \emph{technical} property; we ask the
orthogonal \emph{normative} question of compliance with an external standard,
which shifts the burden from encoding preferences to faithful application of
an authoritative vocabulary under false-positive control (§\ref{sec:niscan}).
Governance research quantifies code-of-conduct
adoption~\cite{tourani2017codeofconduct} and contributor
demographics~\cite{robles2014genderdiff,terrell2017gender,vasilescu2015gender}
without linking either to identifier choices; Eghbal's account of volunteer
maintenance burden~\cite{eghbal2016roadsbridges} grounds our shift-left
reading (§\ref{sec:discussion}). Methodologically we follow repository-mining
studies of normative questions~\cite{he2026fakestars} under
responsible-mining guidance~\cite{kalliamvakou2014promises,munaiah2017curating},
diverging in two ways: the source authority (\ini{}) is external and predates
measurement.
Although \niscan{} shares similar goals of detecting non-inclusive language as several tools such as \code{woke}~\cite{woke},
\code{blocklint}~\cite{blocklint}, GitHubInclusifier's repair
pipeline~\cite{todd2024githubinclusifier}, harmful-term
classifiers~\cite{jacas2025llmharmful,winchester2023harmful} and toxic content detector~\cite{10.1145/3808130}, \niscan{} differs in three key aspects: (1) while existing tools enforce or repair per repository, \niscan{} conducts a empirical study of \allprojects repositories to track ecosystem-wide trends; (2) instead of focusing primarily on Python and Java code identifiers as in GitHubInclusifier~\cite{todd2024githubinclusifier}, \niscan{} provides multi-language support and maps non-inclusive terms across diverse software artifacts like documentation and configurations; and (3) GitHubInclusifier~\cite{todd2024githubinclusifier} uses LLMs as a tool to suggest fixes, we investigate them as a vector for regression, showing how code-generation models reproduce historical non-inclusive naming from their training data. Meanwhile,
Tan et al.~\cite{tan2025harmfulness} measure harm a model \emph{generates};
we measure the residue \emph{already present} in the code such models learn
from.

\section{\niscan{}: A Multilingual Scanner for Non-inclusive Terms}
\label{sec:niscan}

\begin{figure*}[t]
\centering
\includegraphics[width=0.9\textwidth]{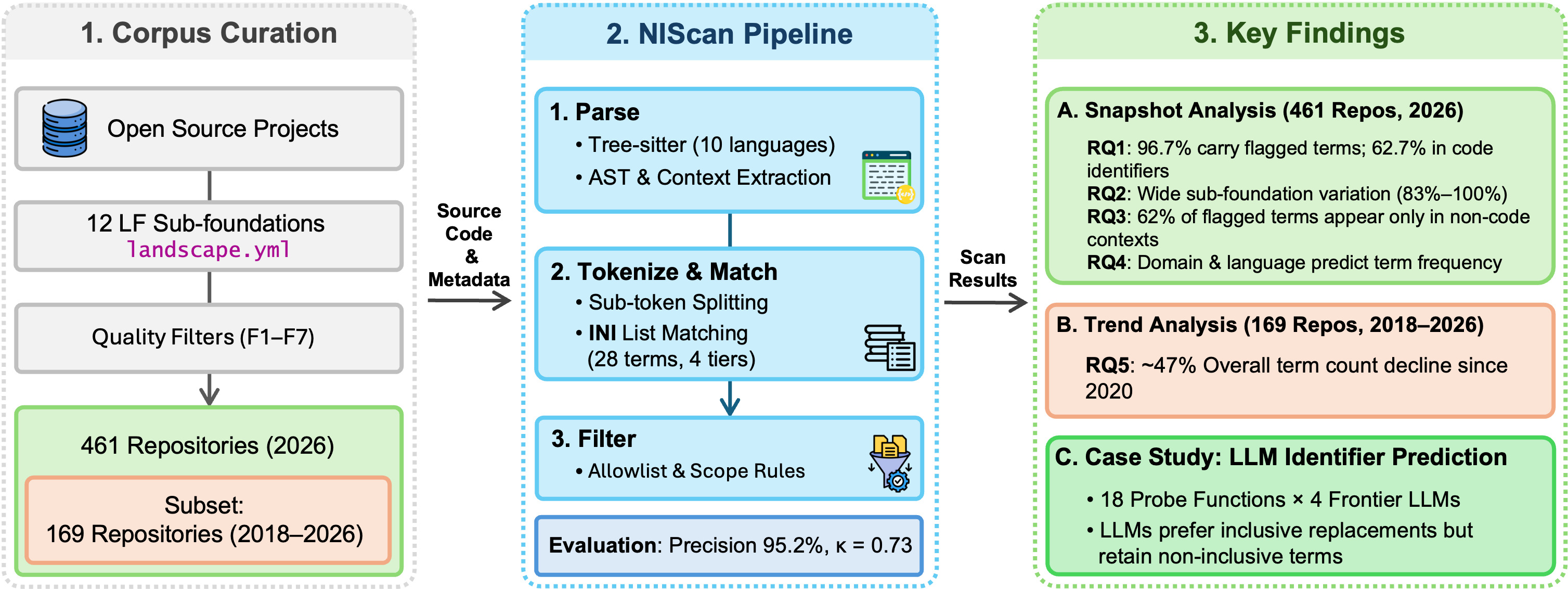}
\caption{Study overview: 461 repositories from 12 LF sub-foundations
$\to$ three-step \niscan{} pipeline $\to$ snapshot analysis (RQ1--RQ4),
trend analysis (RQ5), and an LLM case study.}
\label{fig:overview}
\end{figure*}

\niscan{} is a multilingual scanner designed to support analysis in our empirical study by enabling automated measurement of non-inclusive language across software projects. Specifically, we design \niscan{} according to two criteria: (i) broad language coverage to enable large-scale analysis across heterogeneous codebases, and (ii) produce accurate results with a low rate of false positives so that measurements reflect real-world usage. To satisfy these goals, \niscan{} parses source code in ten programming languages, applies the INI vocabulary without modification, and mitigates false positives through systematic filtering rather than extending or adapting the underlying term set. Rather than introducing a new vocabulary or detection approach, it parses source code in ten programming languages, applies the INI vocabulary without modification, mitigate false positives via filtering.

\subsection{Overview}
\label{sec:niscan:arch}

\niscan{} processes each repository in three stages
(Fig.~\ref{fig:overview}).
\emph{(1)~Parse.} It parses code with tree-sitter and scans documentation
and configuration as plain text, recording every term occurrence with
its \emph{context} (§\ref{sec:niscan:parsing}).
\emph{(2)~Tokenize and match.} 
It normalises identifiers into sub-tokens and tests them against the \ini{} list and its variant table (§\ref{sec:niscan:matching}).
\emph{(3)~Filter.} It uses a rule-based approach to suppress out-of-scope matches.
\niscan{} produces as output a analysis report containing each flagged occurrence with its file, line, term, tier, and context.

Consider the \textsc{etcd} repository\footnote{\url{https://github.com/etcd-io/etcd}}
from our corpus. Its access-control module contains the term \code{whitelist}, which \ini{} flags for imediate replacement. During the \emph{Parse} stage, the tool extracts occurrences from the Go identifier \code{HostWhitelist}, the associated documentation comment, and the configuration key \code{host-whitelist}. In the \emph{Tokenize and Match} stage, the identifier is decomposed into \code{[host, whitelist]}, enabling all three occurrences to be matched against the INI vocabulary. After the \emph{Filter} stage, \niscan{} reports each retained occurrence, together with its tier (i.e., Tier-1 in this example), syntactic context (identifier, comment, and configuration), the files where the term occurs, and their corresponding line numbers.

\paragraph*{Word list and term variants}
\label{sec:niscan:wordlist}

Our vocabulary includes all terms in the \ini{} Word
List~\cite{ini2024wordlist}, which includes 28
terms across four recommendation tiers, from ``adopt immediately''
(Tier-1) to ``no change'' (Tier-0).
\niscan{} flags Tiers~1--3.

\subsection{Parsing}
\label{sec:niscan:parsing}

\niscan{} parses ten languages with tree-sitter~\cite{treesitter}
(C, C++, Go, Java, JavaScript, Python, Ruby, Rust, TypeScript, and
Bash). For each language, it configures grammar-specific node types corresponding to identifiers, comments, and string literals.
Each matched occurrence is classified into one of six syntactic \emph{contexts}: identifier, comment, documentation, string literal, configuration, or filename. Recording context enables subsequent analyses to distinguish where non-inclusive terms appear (e.g., source code, documentation, configuration, or tests). In addition, file paths and line numbers are preserved for every match, allowing occurrences to be localized and attributed to production or test code.

\subsection{Tokenization and matching}
\label{sec:niscan:matching}
\niscan{} splits identifiers into lowercase subtokens using standard naming conventions (e.g., camelCase, PascalCase, snake\_case, and kebab-case). For example, \code{getWhitelistedIPs} is tokenized into \code{[get, whitelisted, ips]}. The tokenizer preserves source offsets so that every reported match includes its exact file, line, and column.

The matcher operates in two modes. For identifiers, it performs \emph{token-exact} matching, requiring a subtoken to exactly match a known variant; consequently, \code{manifest} does not match the term \code{man}. For comments, documentation, string literals, and configuration text, it performs \emph{text} matching using word-boundary regular expressions. Multi-word variants are matched as phrases while allowing flexible whitespace (e.g., \code{white list} matches the canonical term \code{whitelist}).

\subsection{Filtering}
\label{sec:niscan:filtering}
To reduce false positives, \niscan{} applies two filtering rules after matching. First, it uses an \emph{allowlist} derived from the Tier-0 entries of the \ini{} vocabulary. If a matched token appears as part of the allowlist, the match is suppressed. For example, although \code{master} is a Tier-1 term, \code{mastermind} is explicitly classified as acceptable by the \ini{} guidelines, so occurrences of \code{master} within \code{mastermind} are not reported as non-inclusive terms.

Second, \niscan{} applies \emph{exclusion rules} that ignore files and directories unlikely to reflect a project's own naming decisions. These rules exclude third-party dependencies (e.g., \code{vendor/}, \code{node\_modules/}, and \code{third\_party/}), generated source files, software licenses, and changelogs. Such content is typically imported, auto-generated, or historical, and including it would inflate prevalence estimates without representing terminology introduced by project developers.

\subsection{Precision of \niscan{}}
\label{sec:niscan:eval}

To evaluate the precision of \niscan{}, we construct a stratified random
sample of 252 matches from the complete corpus. The sample is stratified
by tier and context, with the number of samples per repository capped to
prevent large repositories from dominating any stratum. The first two authors (annotators),
each with more than four years of software engineering experience,
independently classified every sampled match as either a true positive or
false positive.  To assess annotation reliability, 19\% of the sample was
independently annotated by both annotators, and all disagreements were
resolved through discussion.
The complete sampling
and annotation protocol will be released with our replication package upon acceptance.

Across all sampled matches, \niscan{} achieves a precision of 95.2\%
(Wilson 95\% CI [91.9, 97.3]).  Precision is even higher
for Tier-1 identifier matches, which underpin our primary
analyses, reaching 100\% (62 samples; Wilson 95\% CI [94.2,
100.0]).
Manual inspection shows that the
12 false positives
originate from third-party artifacts (e.g., Kubernetes CRD manifests and upstream Helm chart templates) or externally sourced documentation that should have been excluded from the analysis.
These cases reflect limitations in our artifact filtering rather than incorrect identifier matching. Based on these observations, we extended the exclusion rules to cover additional non-standard vendoring paths.
Inter-rater agreement on the 49
double-annotated samples is substantial ($\kappa = 0.73$).
Overall, our evaluation shows that \niscan{} identifies naming instances with high precision, particularly for the Tier-1 identifier matches used in our primary analyses.

\section{Empirical Analysis of Inclusive Naming in Open-Source Projects}
\label{sec:study}
Our study addresses the following research questions:

\begin{description}[leftmargin=*]
    \item[RQ1:] How prevalent are non-inclusive terms in Linux
Foundation repositories?
    \item[RQ2:] Does foundation governance associate with variation in non-inclusive terminology?
    \item[RQ3:] How are non-inclusive terms distributed between code and non-code context?
    \item[RQ4:] What project characteristics predict non-inclusive terms?
    \item[RQ5:] Have non-inclusive terms declined over time?
\end{description}

\paragraph*{Corpus}
\label{corpus}
A sub-foundation qualifies if it publishes a \code{landscape.yml}
that declares canonical project repositories (§\ref{sec:bg:lf}); we identify twelve that do (CNCF,
LF~AI~\&~Data, LF~Energy, LF~Edge, ASWF, Hyperledger, FINOS,
CD~Foundation, LF~Public~Health, Open~Mainframe, GraphQL, and the
PyTorch Foundation), and include every repository each
landscape lists, making the corpus a \emph{census} of those
sub-foundations rather than a sample. §\ref{sec:threats} gives the full criterion and names six
sub-foundations that fail it. We then apply seven
quality filters (F1--F7) that operationalise ``actively maintained''
following established mining
guidance~\cite{kalliamvakou2014promises,munaiah2017curating}:
$\ge$50 commits, $\ge$50\,KB source, $\ge$5 contributors, pushed
within 24 months, not a fork, documentation-only, or archived. The resulting corpus contains
461 repositories. RQ1--RQ3 draw on all 461; RQ4 re-scans a fixed
cohort of 169 repositories (born $\le$2018) at every year-end.

\subsection{RQ1: Prevalence of Non-inclusive Terms in Linux Foundation Repositories}
\label{sec:rq1}

Among the 461 scanned repositories, 446 (96.7\,\%) contain at least one \ini{}-flagged terms across all tiers and contexts. Restricting the analysis to Tier-1 terms in code identifiers, which \ini{} recommends replacing immediately, 289 (62.7\,\%, CI
[58.2\,\%, 67.0\,\%]) still contain at least one occurrence.
 This shows that \emph{linguistic technical debt persists even for the highest-priority terms} with established replacements.
Only 15 repositories (3.3\%) contain no \ini{}-flagged terms. Table~\ref{tab:findings-by-term} shows the prevalence of each \ini{}-flagged term across the corpus. The \emph{Count} column reports the total number of occurrences of each canonical term, \emph{\%} column denotes its proportion of all matched occurrences
, and \emph{Repos} reports the number of repositories containing at least one occurrence of the term. Overall, the corpus contains 133{,}541 matches, 
a mean of 290 occurrences per repository. 

\begin{finding}[Finding 1: \ini{}-flagged terms remain widespread]
We detect at least one \ini{}-flagged term in 96.7\,\% of repositories,
and 62.7\,\% contain a Tier-1 term in a code identifier. Five years after the guidance, only 3.3\,\%
contain no flagged terms.
\end{finding}

\paragraph*{Term distribution} 
As shown in Table~\ref{tab:findings-by-term}, the distribution is highly skewed toward a small number of terms.
Specifically, \code{master} alone accounts for
64.3\,\% of all matches, and \code{master}~+~\code{abort} together
account for 82.5\,\%.
The distribution exhibits a pronounced long tail: beyond the eight most frequent terms, the remaining nine Tier 1--3 terms collectively account for only 0.9\% of all matches.
These results indicate that most linguistic technical debt is concentrated in a small subset of high-frequency terms.

\begin{table}[t]
\centering
\caption{Matches by canonical \ini{} term (top eight by count).
Repos counts are per-term.}
\label{tab:findings-by-term}
\small
\begin{tabular}{lc rr r}
\toprule
Term & Tier & Count & \% & Repos \\
\midrule
\code{master}       & 1 & 85{,}823 & 64.3 & 431 \\
\code{abort}        & 1 & 24{,}354 & 18.2 & 296 \\
\code{whitelist}    & 1 &  8{,}356 &  6.3 & 200 \\
\code{master-slave} & 1 &  7{,}098 &  5.3 &  97 \\
\code{sanity-check} & 2 &  5{,}217 &  3.9 & 222 \\
\code{man-in-the-middle} & 3 &    577 & 0.4 &  59 \\
\code{end-of-life}  & 3 &      528 &  0.4 &  50 \\
\code{segregate}    & 2 &      436 &  0.3 &  51 \\
remaining 9 terms   & -- &  1{,}152 & 0.9 &  -- \\
\midrule
\textbf{Total}      & -- & \textbf{133{,}541} & \textbf{100.0} & \textbf{446} \\
\bottomrule
\end{tabular}
\end{table}

\begin{finding}[Finding 2: Two terms dominate]
\code{master} and \code{abort} together produce 82.5\,\% of all
\ini{}-flagged matches. A targeted remediation of these two terms
would address four-fifths of the residue corpus-wide.
\end{finding}

\paragraph*{Context distribution}
Each match falls in one of six contexts. By raw count the corpus splits into
31.9\,\% identifier, 28.9\,\% documentation, 15.0\,\% configuration,
14.5\,\% string literal, 9.2\,\% comment, and 0.6\,\% filename, so a majority
(68.1\,\%) of residue sits outside code identifiers. Because this
count is dominated by a few very large repositories, we also measure
the fraction of repositories containing at least one match per context;
this flips the ranking: documentation appears in the most
repositories (87.6\,\%) and identifiers in the fewest (64.4\,\%). A
per-(repository, term) analysis sharpens the point: 386 of the 446
repositories with at least one appearing term (86.5\,\%) carry a flagged
term in comments or documentation but in no code identifier, and across all
1{,}558 (repository, term) pairs, non-code-only instances outnumber
code-only ones 52.6\,\% to 4.6\,\%. A code-only scan or linter therefore
under-reports ecosystem-visible residue.

\paragraph*{Precision}
The 62.7\,\% Tier-1 identifier prevalence counts only Tier-1 terms in
identifier context, which the precision sample (§\ref{sec:niscan:eval})
validates at 100\,\% precision (62 samples, Wilson 95\,\% CI
[94.2, 100.0]; 95.2\,\% across all tiers and contexts). Tier-3 ``be
cautious'' terms and occurrences in strings, configuration, or filenames
are excluded from this figure.

\subsection{RQ2: Association Between Foundation Governance and Non-inclusive Terms in Linux Foundation Repositories}
\label{sec:rq2}

\paragraph*{Motivation}
As Linux Foundation co-founded the \ini{} (§\ref{sec:intro}), its recommendations are expected to propagate most directly within LF repositories. However, the twelve sub-foundations differ in age, domain, and the extent to which naming guidance is reflected in their contribution policies. RQ2 examines whether these differences are associated with variation in the prevalence of non-inclusive terminology across sub-foundations.

\paragraph*{Sub-foundation variation}
Table~\ref{tab:rq2-foundations} reports each sub-foundation's
\emph{has-match rate}: the proportion of its repositories that contain at
least one \ini{}-flagged term in any tier and context.
For each has-match rate, we report a Wilson-score 95,\% confidence interval (CI) to quantify uncertainty in the estimated proportion  where wider intervals indicate less precise estimates resulting from smaller sample sizes.
The $n$ column in Table~\ref{tab:rq2-foundations} denotes
the number of scanned repositories for each sub-foundation, whereas the last column shows the
Wilson-score 95\,\% confidence interval (CI) for each rate. 
Sample sizes range from $n=11$ (Open-Mainframe) and $n=12$ (Hyperledger) to $n=231$ (CNCF).
Overall, all analyzed sub-foundations exhibit high \emph{has-match} rates (83.3--100.0\,\%), including several near-saturation cases ($\ge$97\%), across heterogeneous sample sizes.

\paragraph*{Maturity-level variation within CNCF}
Given that the \emph{has-match} rate is a coarse repository-level metric and several sub-foundations contain small samples ($n < 20$), we conduct a finer-grained analysis of within-sub-foundation structure, focusing on maturity levels. Within CNCF, the maturity hierarchy (i.e., sandbox, incubating, graduated, as explained in §\ref{sec:bg:lf}), shows differences in the average number of matches per repository: graduated projects report 721 matches on average, compared to 182 for sandbox projects. 
In the evaluated dataset, graduated projects show higher numbers of \ini{}-flagged terms compared to other maturity levels. CNCF graduation reflects production readiness criteria rather than naming practices.

\begin{finding}[Finding 3: High prevalence of \ini{}-flagged terms with variation across sub-foundations and maturity levels]
  The proportion of repositories containing at least one flagged term ranges from 83.3\,\% (LF-Energy) to 99.6\,\% (CNCF). Within CNCF, flagged-term counts vary across maturity levels, with graduated projects showing higher average counts per repository than sandbox projects.
\end{finding}

\paragraph*{Absence of formal governance adoption}
We define \emph{formal governance adoption} as explicit repository-level mechanisms enforcing \ini{}-aligned naming, including governance documentation, policy configuration, or automated enforcement (e.g., linters). We searched all 461 repositories for such evidence and explicit references to the \ini{}. Our results reveal that only three repositories contain explicit commitments: two mention the Initiative in \code{CONTRIBUTING.md}, and one includes a grep-based linter~\cite{woke}. A further 14\,\% adopt the Contributor Covenant's ``inclusive language'' clause, which we exclude as it targets communication style rather than identifier naming. Despite the near-absence of formal adoption, flagged-term density decreases by approximately 47\,\% relative to the pre-2020 baseline (RQ4, §\ref{sec:rq4}), consistent with ecosystem-level changes such as the 2020 \code{master}$\rightarrow$\code{main} transition. Even among repositories with explicit policies, enforcement gaps remain. For example, 
one project enables the linter, but legacy terminology such as \code{master} persists in generated release artifacts. 

\begin{table}[t]
\centering
\caption{Proportion of repositories containing at least one
\ini{}-flagged terms, per LF sub-foundation (Wilson 95\,\% CIs);
$n$ is the number of scanned repositories. Sub-foundations with fewer
than 10 repositories (LF-Edge, PyTorch, CD~Foundation, GraphQL) are
omitted as too small for a reliable rate; no LF~Public~Health
repository passed the quality filters (§\ref{sec:study}).}
\label{tab:rq2-foundations}
\small
\begin{tabular}{l r r l}
\toprule
Sub-foundation & $n$ & Match (\%) & 95\,\% CI \\
\midrule
FINOS          &  65 & 100.0\,\% & [94.4, 100.0] \\
Hyperledger    &  12 & 100.0\,\% & [75.8, 100.0] \\
CNCF           & 231 & 99.6\,\% & [97.6, 99.9] \\
LF-AI-Data     &  47 & 97.9\,\% & [88.9, 99.6] \\
Open-Mainframe &  11 & 90.9\,\% & [62.3, 98.4] \\
ASWF           &  17 & 88.2\,\% & [65.7, 96.7] \\
LF-Energy      &  60 & 83.3\,\% & [72.0, 90.7] \\
\bottomrule
\end{tabular}
\end{table}

\subsection{RQ3: Distribution of Non-inclusive Terms Across Code and Non-code Contexts}
\label{sec:rq3}

\paragraph*{Motivation and definitions}
RQ3 investigates how non-inclusive terms are distributed across different contexts within a repository. 
Analyses based solely on source code identifiers may miss how widely such terms persist in the other five contexts \niscan{} tracks (comments, documentation, string literals, configuration files, and filenames), all of which shape developer-facing communication but none of which a code-only scan or linter would
check.

We group these five non-identifier contexts as \emph{non-code} and, for each (repository, term) pair, record whether the term appears in code identifiers only (\emph{code-only}), in non-code contexts only (\emph{non-code-only}), or in \emph{both}. Non-code-only terms are the interesting case: they are absent from the implementation yet present in developer-facing text, exactly what a code-only scan or linter would miss.

\paragraph*{Non-code text dominates}
Of the 1{,}558 (repository, term) pairs in the corpus, 62.3\,\% are
non-code-only, 35.5\,\% appear in both, and only 2.2\,\% are code-only:
when a term occupies one category but not the other, it is the identifier
category that is empty far more often. At the repository level, 416 of
the 446 repositories with at least one match (93.3\,\%) carry a flagged
term in a non-identifier context but in no code identifier.

The split is term-dependent. \code{abort}, a C/POSIX library function
that is naturally an identifier, is non-code-only in 35.5\,\% of its
(repository, term) pairs, whereas conceptual Tier-3 terms such as
\code{blast-radius} and \code{end-of-life} are non-code-only in 100\,\%
of cases. The intermediate cases are Tier-1 terms that can be
identifiers but often are not: \code{master} is non-code-only in
56.4\,\% of its (repository, term) pairs and \code{whitelist} in
51.0\,\%.

\paragraph*{Example}
In etcd, a graduated CNCF key-value store, \code{whitelist} spans
three contexts: the Go field \code{HostWhitelist} and test function
\code{TestIsHostWhitelisted} (identifier), comments documenting them,
and string literals such as the flag help text ``white list of origins.''
A scanner that checks only identifiers would miss the comments and
strings entirely.

\begin{finding}[Finding 4: Most \ini{}-flagged terms appears in non-identifier contexts]
In 93.3\,\% of repositories at least one flagged term appears in a
non-identifier context (comments, documentation, strings, configuration,
or filenames) but in no code identifier, and non-code-only (repository,
term) pairs outnumber code-only ones 62.3\,\% to 2.2\,\%. A code-only
scan therefore under-reports non-inclusive terms.
\end{finding}

\subsection{RQ4: What project characteristics predict non-inclusive terms?}
\label{sec:rq4}

\paragraph*{Motivation and model}
RQ1--RQ3 establish the prevalence, governance dimension, and context
distribution of non-inclusive terms; this RQ asks which projects carry
them. Repository size is already characterised (it dominates the raw
counts, §\ref{sec:rq2}), so we retain it as a control and turn to a
characteristic the previous RQs do not explain: what a project does. We
classify each repository into one of six \emph{functionality domains}
with a transparent keyword map over its GitHub description and topics
(18\,\% fall in a residual ``Other'' class), and fit a
negative-binomial regression of the per-repository identifier-context
match count on domain, primary language, size, activity, and popularity.
Table~\ref{tab:rq4-model} reports \emph{incidence-rate ratios}
(IRR~$= e^{\beta}$): for categorical predictors, the IRR is the ratio
of the expected count to the reference category, holding all other
predictors fixed. An IRR of~5 means five times as many flagged
identifiers as the reference.

\begin{table}[t]
\centering
\caption{Negative-binomial regression of per-repository identifier-context
match count ($n = 461$).
IRR $= e^{\beta}$: the multiplicative change in the expected count per
unit change in the predictor; IRR ${>}\,1$ $\Rightarrow$ more flagged terms.
Reference: domain ``Other'', language C++ (highest).
Numeric predictors log-transformed; language coefficients in the text.}
\label{tab:rq4-model}
\small
\begin{tabular}{l r l r}
\toprule
Predictor & IRR & 95\,\% CI & $p$ \\
\midrule
\multicolumn{4}{@{}l}{\emph{Functionality domain} (vs.\ Other):}\\
\quad ML \& data             & 7.29 & [4.56, 11.64] & $<10^{-16}$ \\
\quad Networking/systems     & 5.21 & [2.90, 9.38]  & $<10^{-7}$  \\
\quad Infrastructure/orchestration & 5.03 & [3.31, 7.66] & $<10^{-13}$ \\
\quad Web/application        & 3.74 & [1.57, 8.90]  & 0.003       \\
\quad Observability/security & 3.39 & [1.70, 6.77]  & $<10^{-3}$  \\
\quad Developer tools/libs   & 0.37 & [0.18, 0.73]  & 0.004       \\
\midrule
\multicolumn{4}{@{}l}{\emph{Controls}:}\\
\quad $\log$ size            & 1.82 & [1.64, 2.01] & $<10^{-30}$ \\
\quad $\log$ contributors    & 1.24 & [1.03, 1.49] & 0.021       \\
\quad $\log$ stars           & 1.11 & [1.01, 1.22] & 0.035       \\
\quad $\log$ commits         & 0.87 & [0.74, 1.02] & 0.092       \\
\bottomrule
\end{tabular}
\end{table}

\begin{figure}[t]
\centering
\includegraphics[width=\columnwidth]{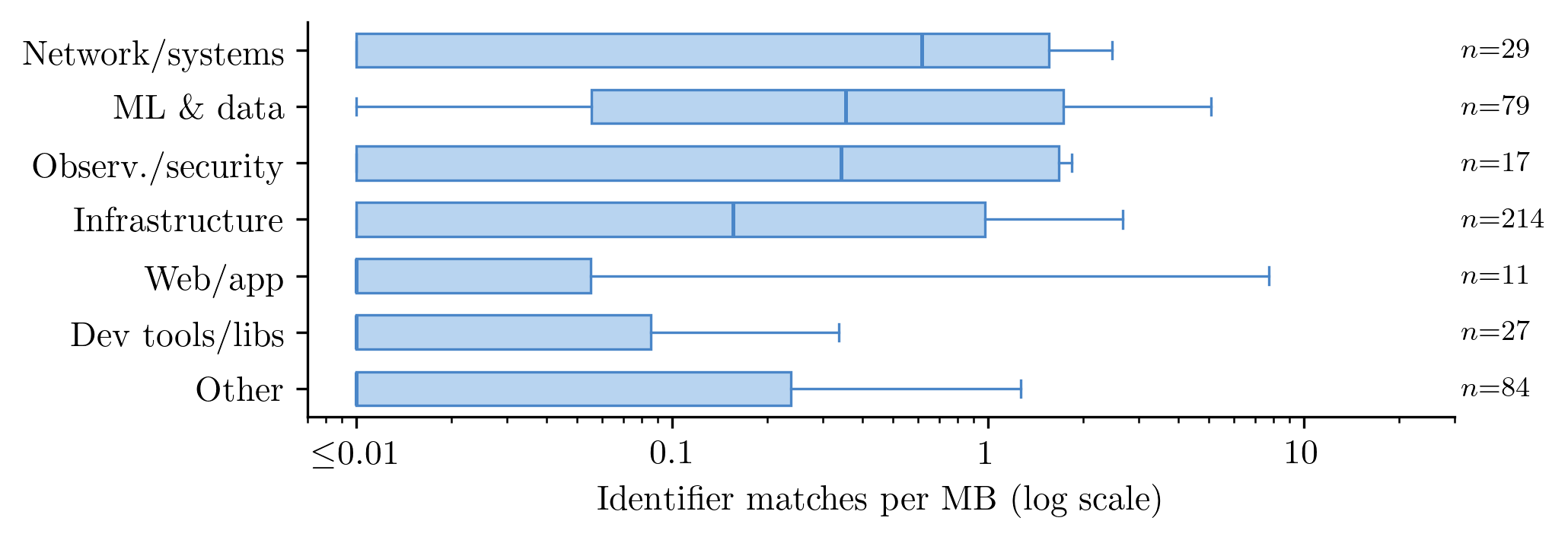}
\caption{Per-repository identifier density by functionality domain (log scale,
zeros clamped to the floor; $n$ per box at right). Boxes span the interquartile
range with the median marked; whiskers span the 10th--90th percentiles.
Infrastructure-class domains carry more flagged terms per
MB than developer-tool and Other projects; the regression
(Table~\ref{tab:rq4-model}) confirms the gap controlling for language and size.}
\label{fig:rq4-density}
\end{figure}

\paragraph*{Functionality is the strongest controllable predictor}
Five of the six classified domains carry significantly more flagged terms
than general projects, holding size, language, and popularity fixed: ML
\& data (IRR~7.3), networking/systems (5.2),
infrastructure/orchestration (5.0), web/application (3.7), and
observability/security (3.4) each multiply the expected
identifier-context match count by 3.4--7.3$\times$ relative to the
``Other'' baseline (all $p < 0.003$). Developer-tool/library projects
carry significantly \emph{fewer} flagged terms (IRR~0.37, $p = 0.004$),
consistent with these projects having fewer infrastructure-API
dependencies. The raw density agrees
(Fig.~\ref{fig:rq4-density}): the median identifier matches per MB
range from 0.16 to 0.62 for the four infrastructure-class domains,
against 0 for developer-tool, web, and Other projects. The mechanism is
semantic, not stylistic: the dominant \ini{} terms name infrastructure
concepts directly, \code{master}/\code{slave} in replication and
orchestration, \code{whitelist}/\code{blacklist} in networking and
security, \code{abort} in systems code, so the domains built around
those concepts inherit the vocabulary. The effect survives controlling
for language, so it is not a restatement of ``systems languages carry
more'': what a project does predicts its non-inclusive term count
independently of how it is written.

\paragraph*{Language remains a strong independent predictor}
Of the ten parsed languages (§\ref{sec:niscan:parsing}), six have
$\ge$20 repositories and are estimated separately; C, JavaScript, and
the remaining languages are pooled into Other. C++ is the regression
reference (highest estimated count); all other language coefficients are
significant at $p < 0.01$ except Java ($p = 0.052$). Measured as
identifier density (matches per MB, which removes the size effect by
construction), Rust leads at a median of 1.3, followed by C++ (0.6);
Go and the managed languages range from 0.09 to 0.15. The likely
mechanism is again specific: \code{abort()} is a
\code{libc}/\code{<cstdlib>} function inherited as an unavoidable API
name.

\paragraph*{Flagged-term composition is ecosystem-specific}
Beyond how many flagged terms a language carries, the package ecosystem
shapes which terms dominate. Identifier-context matches in \code{npm}
(JavaScript and TypeScript) and \code{Cargo} (Rust) projects are
overwhelmingly \code{abort} (73\,\% and 71\,\% of their flagged
identifiers), inherited from \code{AbortController} and Rust's
\code{abort}; Java/Maven matches are disproportionately the
\code{master}--\code{slave} compound (27\,\%, against $\le$4\,\% in
every other ecosystem), the vocabulary of its replication and CI
conventions; and Go and Python projects lead with \code{master}
(64\,\% and 65\,\%). Each ecosystem reproduces the non-inclusive terms
baked into its core APIs, consistent with terminology diffusing along
ecosystem lines rather than being chosen afresh per project.

\paragraph*{Robustness: sub-foundation grouping}
As a robustness check, we replaced the keyword-derived domain with the
externally defined sub-foundation grouping (merging the four
sub-foundations with fewer than ten repositories). The pattern is
directionally consistent: LF~AI~\&~Data, whose portfolio is
predominantly ML/data-pipeline projects, carries the most flagged
identifiers (IRR~2.38 vs.\ CNCF, $p < 10^{-3}$), while
Open~Mainframe carries the fewest (IRR~0.09, $p < 10^{-4}$),
confirming that the domain effect is not an artefact of our
classification.

\paragraph*{What does not predict non-inclusive terms}
Project age (creation era) shows a weak bivariate correlation with
density (Spearman $\rho = -0.12$, $p = 0.008$) but the regression bins
are non-monotone: the oldest cohort (pre-2014, $n = 17$) is marginally
elevated (IRR~1.96, $p = 0.057$), the 2017--2019 cohort is
\emph{below} the 2020+ reference (IRR~0.44, $p < 10^{-7}$), and the
2014--2016 bin is not significant. No monotone age gradient emerges.
Non-inclusive term counts track what a project does and how it is built,
not when it began.

\begin{finding}[Finding 5: Functionality, ecosystem, and language predict
  non-inclusive term counts]
Controlling for size, language, and popularity, a project's functionality
domain is the strongest controllable predictor of identifier-context
non-inclusive terms: five domains carry 3.4--7.3$\times$ the flagged
terms of general projects (all $p < 0.003$), while developer-tool/library
projects carry significantly fewer (IRR~0.37). Flagged-term composition
is further ecosystem-specific (e.g.\ \code{abort} in npm and Cargo,
\code{master}--\code{slave} in Java), and language is independently
strong (C++ and Rust densest). Project age is not robust.
\end{finding}

\subsection{RQ5: Have non-inclusive terms declined over time?}
\label{sec:rq5}

\paragraph*{Motivation}
Have non-inclusive terms actually declined since the 2020 GitHub
\code{master}$\rightarrow$\code{main}
rename~\cite{github2020main} and the 2020--2021 \ini{}
launch~\cite{ini2024wordlist}? The obvious approach: searching each
term's git history for its first introduction (\code{git log -S}),
suffers from survivorship bias: it only sees terms that are still
present today, so the apparent introduction rate rises
toward the present. The bias is severe: the pickaxe reports a
2.7$\times$ increase on our corpus, while our year-end re-scan
(below) shows a $\sim$47\,\% decline, reversing the direction of the
trend entirely.

\paragraph*{Year-end re-scan design}
We instead re-scan each repository's full tree at every year-end. We
fix a cohort of 169 LF repositories born on or before 2018 so that
each has a pre-initiative baseline. For every year~$Y$, we check out
the last commit before December~31
(\code{git rev-list -1 -{}-before=$Y$-12-31}) and run \niscan{} on
that snapshot. Because each measurement captures the actual tree at
time~$Y$, it reflects both introductions and removals, not just
surviving terms.

\paragraph*{Normalising for code growth}
Flagged-term counts scale sub-linearly with code size
(§\ref{sec:rq4}), so any per-byte or per-KLOC density drops as a
codebase grows even if no terms are added or removed. We therefore
report \emph{matches per 1{,}000 identifiers}: identifiers are the
naming decisions a developer could have made non-inclusively, making
them a more stable baseline. \niscan{}'s extractor counts both
flagged terms and total identifiers over the same file set (vendored,
generated, and license files excluded); we confirm the trend holds
under per-KLOC and per-MB as well. We compare pre~($\le$2019)
against post~($\ge$2022), excluding the 2020--2021 transition period,
which mixes the GitHub rename, the \ini{} launch, and broader 2020
events.

\begin{figure}[t]
\centering
\includegraphics[width=\columnwidth]{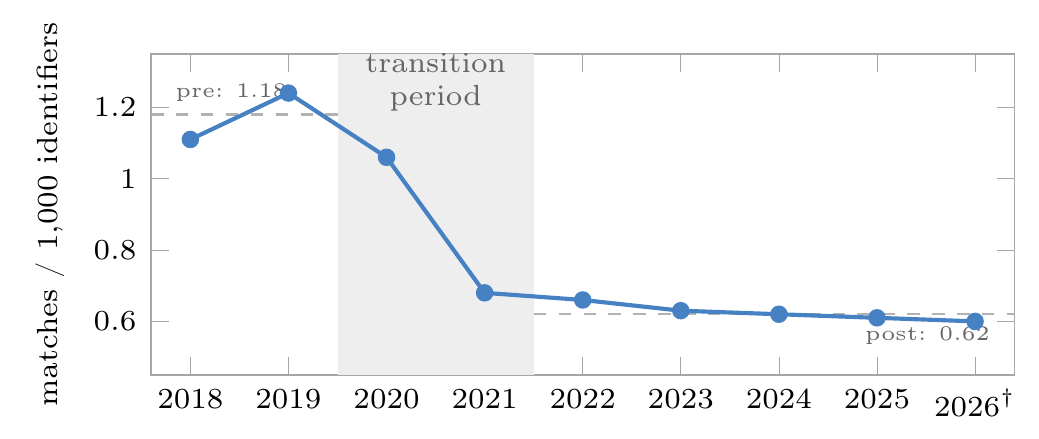}
\caption{Year-end re-scan term-density trajectory (169 LF
repositories born $\le$2018); shaded band: 2020--2021 transition
period (excluded from pre/post comparison). Per-KLOC/per-MB trajectories are parallel (0.52, 0.50).
$^\dagger$2026 partial.}
\label{fig:rq5-trajectory}
\end{figure}

\paragraph*{Non-inclusive terms have declined by $\sim$47\,\%}
Pooled across the cohort, per-1k-identifier density drops from
1.18 (pre) to 0.62 (post), a $\sim$47\,\% decline. The result is
consistent across denominators (per-KLOC ratio 0.52, per-MB 0.50).
The decline is not driven by a few large projects: 125 of 169
repositories (74.0\,\%) individually show a decrease (Wilcoxon
$p = 9.8\times10^{-9}$).

\paragraph*{A sharp drop, then a plateau}
The trajectory (Fig.~\ref{fig:rq5-trajectory}) is not a steady decline.
Density peaks in 2019, drops sharply during 2020--2021, and then
stabilizes at approximately 0.60 for the next five years.
This decline temporally aligns with the GitHub rename and the lead-up to \ini{},
while the plateau begins after \ini{}'s formal training program started in late 2021.
One plausible interpretation is that easily removable terms were addressed during this
two-year window, after which further progress slowed.
However, we cannot determine whether the plateau reflects exhaustion of easily renamed
terms or diminishing intervention impact, because the GitHub rename, the \ini{} launch,
and broader 2020 events overlap in time.

\begin{finding}[Finding 6: Non-inclusive term prevalence has declined
  by $\sim$47\,\%, reversing the git-history trend]
Year-end re-scans of each repository's full tree show that
per-1k-identifier term density fell from 1.18 to 0.62 ($\sim$47\,\%
decline, consistent under per-KLOC and per-MB; 74.0\,\% of
repositories declined, Wilcoxon $p=9.8\times10^{-9}$). A
\code{git log~-S} analysis on the same data reports a 2.7$\times$
\emph{increase}, illustrating how survivorship bias can invert the
trend. The decline is a sharp drop across 2020--2021 followed by a
five-year plateau at half the peak, and is driven by terms that
require manual renaming (93.2\,\% of the post-period total), not by
the automated \code{master}$\rightarrow$\code{main} branch rename.
\end{finding}

\subsection{Case Study: LLM Reconstruction of Non-Inclusive Terms}
\label{sec:discussion:probe}
\paragraph*{Motivation}
While RQ1--RQ5 characterize the prevalence, distribution, and evolution of non-inclusive terms in Linux Foundation (LF) repositories, they do not reveal whether the remaining naming residue has downstream consequences. 
To bridge this gap, we frame our investigation around the task of automated method name prediction~\cite{allamanis2015suggesting,wang2023pre,qian2024exploring}. 
which studies how method identifiers can be inferred from surrounding program context. 
The focus on function-level identifiers is motivated by both practical and theoretical considerations. Practically, large-scale refactoring efforts in systems such as CNCF, which actively rename non-inclusive function identifiers, highlight the real-world engineering relevance of naming conventions and their propagation through code evolution. Theoretically, prior work has shown that function names are important in program comprehension and are strongly coupled with the semantics of the underlying implementation~\cite{allamanis2015suggesting}. This makes function-level identifiers a well-studied and appropriate setting to study if contextual cues alone are sufficient to recover or reintroduce legacy terminology. As
large 
language models are increasingly used for code completion and generation, such residual cues may further lead models to reconstruct non-inclusive identifiers even after they have been renamed.

\paragraph*{Probe-set construction}
We apply a three-step selection pipeline.
\emph{Step~1 (identification):}
\niscan{} identifies every function or method definition in the
corpus (§\ref{corpus}) whose name contains any Tier-1 \ini{}
term, yielding
2{,}371 candidate functions across 241~repositories.
\emph{Step~2 (filtering):}
We fetch each candidate's source via the GitHub API, mask every
occurrence of the target term, and retain only functions whose masked
body contains control flow and enough distinct identifiers to make
the prediction task meaningful.
This excludes trivial wrappers (e.g., single-return getters),
bodiless declarations, vendored SDK code, and overly long functions.
\emph{Step~3 (selection):}
We select up to five candidates per term, preferring more complex
bodies, constrained to at most two functions per repository and no
duplicate function names.
Four Tier-1 terms appear frequently enough in function names to
supply candidates: five \code{master}, four \code{master-slave},
five \code{whitelist}, and four \code{abort} (positive controls);
the remaining four (\code{tribe}, \code{blackhat-whitehat},
\code{grandfathered}, \code{cripple}) rarely or never appear in
function names.
The final set comprises 18~functions.

For each function, we mask every occurrence of the target term
and prompt four frontier language models to predict the masked
word from the remaining source code alone.
We select the flagship model from each of the four major providers
as of June~2026: GPT-5.5~\cite{openai2026gpt55},
Claude Opus~4.8~\cite{anthropic2026opus48},
Gemini~3.1~Pro~\cite{google2026gemini31}, and
DeepSeek~V4~Pro~\cite{deepseek2026v4}.
All models are queried via API at medium reasoning effort with a
fixed prompt that does not mention inclusive naming.

The prompt asks
each model to return its top three guesses for the masked identifier,
ranked by confidence; we refer to the first guess as \emph{top-1} and
to any match within the three guesses as \emph{top-3}.

We count a prediction as a reproduction only if it returns the
original non-inclusive term (\code{master}/\code{slave} or
\code{whitelist}/\code{blacklist}), not an inclusive alternative such
as \code{allowlist} or \code{main}.
We report the four \code{abort} functions separately.
Unlike the other terms, \code{abort} names a C standard-library call
with no inclusive replacement; high reproduction in these cases confirm that the probe
reliably recovers a term when the surrounding context fully
determines it. 

\begin{table}[t]
\centering
\caption{Reproduction rate (\%) of the masked non-inclusive term.
  \emph{All}: all 18~probe functions; \emph{\code{abort}}: the
  four \code{abort} functions only.}
\label{tab:probe-rates}
\footnotesize
\setlength{\tabcolsep}{4pt}
\begin{tabular}{lccccc}
\toprule
 & \multicolumn{2}{c}{Top-1} & & \multicolumn{2}{c}{Top-3} \\
\cmidrule(lr){2-3}\cmidrule(lr){5-6}
Model & All & \code{abort} & & All & \code{abort} \\
\midrule
GPT-5.5          & 39 &  75 & & 78 & 100 \\
Opus 4.8         & 50 & 100 & & 78 & 100 \\
Gemini 3.1 Pro   & 50 & 100 & & 83 & 100 \\
DeepSeek V4 Pro  & 22 &  75 & & 61 & 100 \\
\bottomrule
\end{tabular}
\end{table}

\begin{table}[t]
\centering
\caption{Per-function top-1 reproduction: number of models (out of~4)
  that return the masked term as their top prediction.
  $^\dagger$~positive control.}
\label{tab:probe-funcs}
\scriptsize
\setlength{\tabcolsep}{3.5pt}
\resizebox{0.95\columnwidth}{!}{%
\begin{tabular}{llc}
\toprule
Term & Function (repo) & \#/4 \\
\midrule
\multirow{5}{*}{\code{master}}
 & \code{read\_master\_key\_settings} (DESFire) & 2 \\
 & \code{\_find\_latest\_master\_experiment}    & 0 \\
 & \code{rotate\_...\_master\_key} (litellm)    & 1 \\
 & \code{resolve\_master\_key} (vault)          & 0 \\
 & \code{getNearestRemoteMaster...} (IntelliJ)  & 0 \\
\midrule
\multirow{4}{*}{\code{master-slave}}
 & \code{i2c\_slave\_init\_helper}              & 4 \\
 & \code{deinstallSlave} (M-Bus)               & 3 \\
 & \code{assignReplicationSlaves}              & 0 \\
 & \code{onewire\_slave\_wait...}              & 0 \\
\midrule
\multirow{5}{*}{\code{whitelist}}
 & \code{SubjectWhitelisted}                   & 3 \\
 & \code{BlacklistToken} (auth)                & 2 \\
 & \code{createBlacklist}                      & 0 \\
 & \code{IsHostWhitelisted} (etcd)             & 0 \\
 & \code{configure\_ssrf\_whitelist}           & 0 \\
\midrule
\multirow{4}{*}{\code{abort}$^\dagger$}
 & \code{http\_close\_or\_abort\_conn}         & 4 \\
 & \code{DeathTestAbort}                       & 4 \\
 & \code{\_abort\_if\_submit\_incomplete}      & 3 \\
 & \code{assertAbort}                          & 4 \\
\bottomrule
\end{tabular}
}
\end{table}

Table~\ref{tab:probe-rates} summarises the reproduction rate per model,
and Table~\ref{tab:probe-funcs} breaks it down by function.
We observe four patterns.
First, the \code{abort} functions reproduce at high rates
(top-1: 75--100\,\%, top-3: 100\,\%), validating the probe as a
positive control: when the surrounding code fully determines the
identifier, models recover it reliably.
Second, for the remaining 14~functions, models seldom place the
non-inclusive term first but often include it in their top-3,
producing a consistent gap between top-1 and top-3 reproduction.
Third, models frequently predict inclusive alternatives at top-1: for
\code{whitelist} functions they return \code{allowlist} or
\code{allow} (e.g., \code{IsHostWhitelisted}: 0/4,
\code{configure\_ssrf\_whitelist}: 0/4); for \code{master} they
return \code{main}, \code{encryption}, or \code{vault} depending on
context.
Fourth, reproduction is highest where hardware protocols leave no
neutral synonym (I2C \code{slave}: 4/4, M-Bus
\code{deinstallSlave}: 3/4) and lowest where an inclusive
alternative exists (\code{assignReplicationSlaves}: 0/4,
\code{getNearestRemoteMaster}: 0/4).

\begin{finding}[Finding 7: Models down-rank but do not eliminate
  non-inclusive identifiers]
Frontier language models rarely rank non-inclusive identifiers as their
top prediction but frequently retain them among their top three
suggestions. Top-1 reproduction ranges from 22\,\% to 50\,\%, while
top-3 reaches 61\,\%--83\,\%. Models down-rank rather than eliminate
non-inclusive terms, preferring inclusive variants unless the surrounding
code uniquely determines the original identifier.
\end{finding}

\section{Discussions and Implications}
\label{sec:discussion}

We discuss the implications and actionable guidance for maintainers, foundations, and tool builders.

\subsection{Policy is widely adopted but inconsistently reflected in practice}
\label{sec:discussion:policy}
Across the LF foundations, inclusive naming guidance is widely disseminated, yet its adoption remains uneven in practice.
Five years after its introduction, non-inclusive identifiers are still prevalent in LF repositories (only 3.3\% 
repositories are free of non-inclusive terms).
Finding~5 attributed the prevalence to the size of the repository, functionality domain, and primary language: larger
repositories contain more flagged terms, five of six functionality
domains carry 3--7$\times$ more than the baseline, and C/C++ codebases
inherit non-inclusive API names such as \code{abort}.  
Particularly, large C codebases tend to accumulate non-inclusive terms that is not immediately eliminated by governance or policy adoption alone. 

\paragraph*{Implications for LF Repository Maintainers: Governance Does Not Guarantee Practice Alignment}
Finding~6 
suggests that remediation of non-inclusive terminology in LF repositories has largely occurred without formal, per-project governance adoption. Explicit documentation of inclusive-{naming} policies is rare (3 of 461 repositories, 0.65\%),
indicating that maintainers generally do not rely on governance artifacts to drive change. Instead, repository-level transformation appears to be primarily driven by ecosystem-wide defaults, particularly the 2020 \code{master}$\rightarrow$\code{main} transition and the adoption of \code{main} as the default across hosting platforms, templates, and developer tooling. For LF repository maintainers, this implies that governance documentation alone is neither a necessary nor a sufficient mechanism to drive terminology change. Instead, maintainers often inherit naming conventions through infrastructure-level updates rather than explicitly adopting policies. However, this also creates a gap between intended governance and actual repository state: even when explicit policies or linters are enabled, they may not fully propagate across build systems, release pipelines, or generated artifacts, as evidenced by residual occurrences such as \code{master} in release outputs.

Meanwhile, the longitudinal analysis (RQ5) reveals a complementary
dynamic that is not visible in static comparisons. The number of
flagged terms per 1{,}000 identifiers fell $\sim$47\,\% from the
pre-period ($\le$2019) to the post-period ($\ge$2022), driven
predominantly by discretionary terms (93.2\,\% of the post-period
total) rather than by the automated branch rename alone (Finding~6).
Critically, the decline is a step-change in 2020--2021, followed by a
five-year floor: easily removable terms were cleared quickly, but the
remaining non-inclusive vocabulary has not declined further. This
suggests that governance is not the primary driver of non-inclusive terminology
reduction. Instead, ecosystem-wide infrastructure changes (the GitHub
default branch rename, platform templates) drove most
improvement, while the persistent floor reflects terms that are too
deeply embedded in APIs and legacy code for policy alone to eliminate.

\noindent\textbf{Actionable Guidance.} First, policy adoption should be complemented with systematic refactoring efforts, as governance alone does not remove historical naming in large codebases. Second, evaluations of inclusivity interventions should account for temporal change rather than relying on snapshot measurements, since their effects may primarily manifest as differences in cleanup trajectories over time. Third, remediation efforts should prioritize large repositories, where accumulated legacy terminology persists regardless of organizational context.

\subsection{Term Inclusiveness in the context of AI-assisted Coding}
\label{sec:discussion:shiftleft}

Finding~3, 5, and 6 
indicates that addressing non-inclusive terminology early in the development lifecycle is substantially more cost-effective than correcting it after widespread propagation in large codebases, extending prior evidence from testing, security, and accessibility to inclusive naming at ecosystem scale. 
The observed pattern is stable across governance conditions, suggesting it reflects intrinsic development dynamics rather than specific interventions. 
These results also have implications for LLM-assisted and agentic software development, as models are trained on large public repositories in which 96.7\% 
contain flagged terminology, creating a strong prior that may be reproduced in generated code and raising concerns about the provenance of outputs~\cite{xu2026makescodegenerationethically}. The post-2021 stabilization in Fig.~\ref{fig:rq5-trajectory} 
aligns with the adoption of code completion tools,
indicating a potential feedback loop between generated and existing code distributions that warrants empirical validation. Overall, the findings position inclusive naming as a property that must be managed throughout both human and model-driven code production pipelines, with quantitative distributions offering a basis for evaluation.

\noindent\textbf{Actionable Guidance.} First, inclusive naming controls should be embedded at code generation and commit time, as preventing introduction of non-inclusive terminology is more effective than correcting it after accumulation in large codebases. Second, LLM-assisted and agentic development workflows should incorporate automated, deterministic scanning of structured artifacts as a runtime safeguard, since model outputs may reproduce patterns present in training corpora.

\section{Threats to Validity}
\label{sec:threats}

\noindent\textbf{Construct validity.}
To study term inclusiveness defined by the Inclusive Naming Initiative, we use the 28-term \ini{} list without any modification, intentionally avoid extending or editorially modifying the vocabulary because modifications may shift the study from measurement to normative intervention; consequently, organization-specific terms, unmodeled patterns, and model-generated harms~\cite{tan2025harmfulness} remain out of scope by design.

\noindent\textbf{Internal validity.}
We rely on each foundation’s \code{landscape.yml} to define project membership. We initially piloted harvesting full GitHub organizations, but excluded this approach because it introduces unrelated tutorial and product repositories. Our implementation may have bugs that
can affect our results. To mitigate this, we make our
dataset, source code and scripts publicly available.

\noindent\textbf{External validity.}
Our findings generalize to LF repositories that belong to sub-foundations publishing canonical repository landscapes. We select sub-foundation by including all LF
sub-foundations that publishes a canonical-repository
landscape~\cite{cncf2025landscape,nagappan2013diversity}. However, projects hosted directly by LF or by sub-foundations that do not publish such landscapes are outside our scope. Consequently, our results may not be generalized to all open-source projects or to ecosystems with substantially different governance structures.
Projects hosted directly by LF, such as the Linux kernel, do not publish such landscape files and are therefore out of scope by construction.

\noindent\textbf{Conclusion.} Conclusion threats of our evaluation
include subjectivity of manual analysis. We mitigate
the subjectivity of manual analysis by having two annotators
independently labeled each match,
and meet to resolve any disagreement.

\section{Conclusions}
We presented the first ecosystem-scale study of inclusive naming in the Linux Foundation, analyzing \allrepositories{} repositories and 133{,}541 matched occurrences. Our study revealed that non-inclusive terminology remains prevalent (96.7\% of repositories) but has declined by $\sim$47\% since 2019, primarily through ecosystem-wide evolution rather than formal governance. The remaining terms are concentrated outside code identifiers, where code-only scanners are ineffective.
Our case study on four LLMs shows that LLMs still reconstruct them from context (top-3:
61--83\,\%). These results position inclusive naming as a joint
software-evolution, governance, and AI challenge, and our tool and replication data provide a quantitative baseline for future interventions. 
Our artifacts will be made publicly available upon acceptance of this paper.

\bibliographystyle{IEEEtran}
\bibliography{references} 

\end{document}